\documentclass[
    ,final            
  ]
  {aipproc}

\layoutstyle{6x9}

\begin{document}

\title{Beam Modulation System for the Q$_{weak}$ Experiment at Jefferson Lab}


\author{Nuruzzaman (Q$_{weak}$ Collaboration)}{
  address={Hampton University, Hampton, VA 23668, USA.}
}
\begin{abstract}
The Q$_{weak}$ experiment aims to measure the weak charge of the proton to 4\% via the parity-violating asymmetry ($\sim$-200 ppb) in elastic e+p scattering. The scattering rate is largely influenced by five beam parameters: horizontal position (X), angle (X$^{\prime}$), vertical position (Y), angle (Y$^{\prime}$), and beam energy (E). Changes in these parameters when the beam polarization is reversed will create a false asymmetry. To measure the detector sensitivities, we modulate X, X$^{\prime}$, Y, Y$^{\prime}$ using four air-core dipoles in the Hall C beamline as well as beam energy with a superconducting RF cavity. Linear regression is then used to remove the false asymmetry. The beam modulation system has also proven valuable for tracking changes in the Hall-C beamline optics including dispersion at the target.

\end{abstract}

\maketitle



\section{Introduction}

The Standard Model has been successful for describing most elementary particle data  \cite{Novaes:1999} although the model is known to be incomplete. The objective of the Q$_{weak}$ experiment is to search for new parity-violating physics at the TeV scale through a 4$\%$ measurement of the weak charge of proton \cite{website:qweak}. The measured parity-violating asymmetry can be expressed as the difference over the sum of cross sections of different helicity states [eq\eqref{asym} left].
The $\frac{\partial A}{\partial T_{i}}$ are detector sensitivities and the $\Delta T_{i}$ are helicity correlated beam parameter differences [eq\eqref{asym} right]:

\begin{equation}
A_{measured} = \frac{\sigma^{+}-\sigma^{-}}{\sigma^{+}+\sigma^{-}}; \hspace{1.0cm} 
A_{PHYSICS} \equiv A_{measured} - \sum_{i=X,X^{\prime},Y,Y^{\prime},E}{} \left( \frac{\partial A}{\partial T_{i}} \right)  \Delta T_{i}
\label{asym}
\end{equation}

A challenge for the Q$_{weak}$ experiment was to keep these helicity-correlated parameter differences as small as possible and measure the detector sensitivities.



\section{Beam Modulation System}
\label{Beam Modulation System}
Two air-core dipoles separated by $\sim$10 m in the Hall-C beamline were driven simultaneously to get desired position and angle changes at the target. 
The motivations for using a pair over a single dipole were greater flexibility in coil positioning, coil positions being independent of beamline optics, and a compact configuration.
A programmable function generator (VME-4145) was used to produce a sinusoidal signal of 125 Hz which was  sent to a power amplifier (JLab TRIM-I) and then to the air-core dipoles. Sinusoidal waves were used because bench tests showed that JLab TRIM-I could not drive suitable square waves.
Readback signals from the function generator, the power amplifiers, and current transducers (LEM CT 10-T) were sent to  analog to digital converters (ADC). The current transducers were disconnected during production running as they induced a small motion in the beam.

Each beam parameter X, X$^{\prime}$, Y, Y$^{\prime}$, or E was driven for 510 cycles ($\sim$4 s) concurrently with normal data taking. Due to network and other reconfiguration overhead, there was a $\sim$50 s gap between each parameter and 
a complete cycle through all parameters took $\sim$4 min.


%


\section{Analysis}

\begin{figure}[th]
  \includegraphics[height=0.42\textheight]{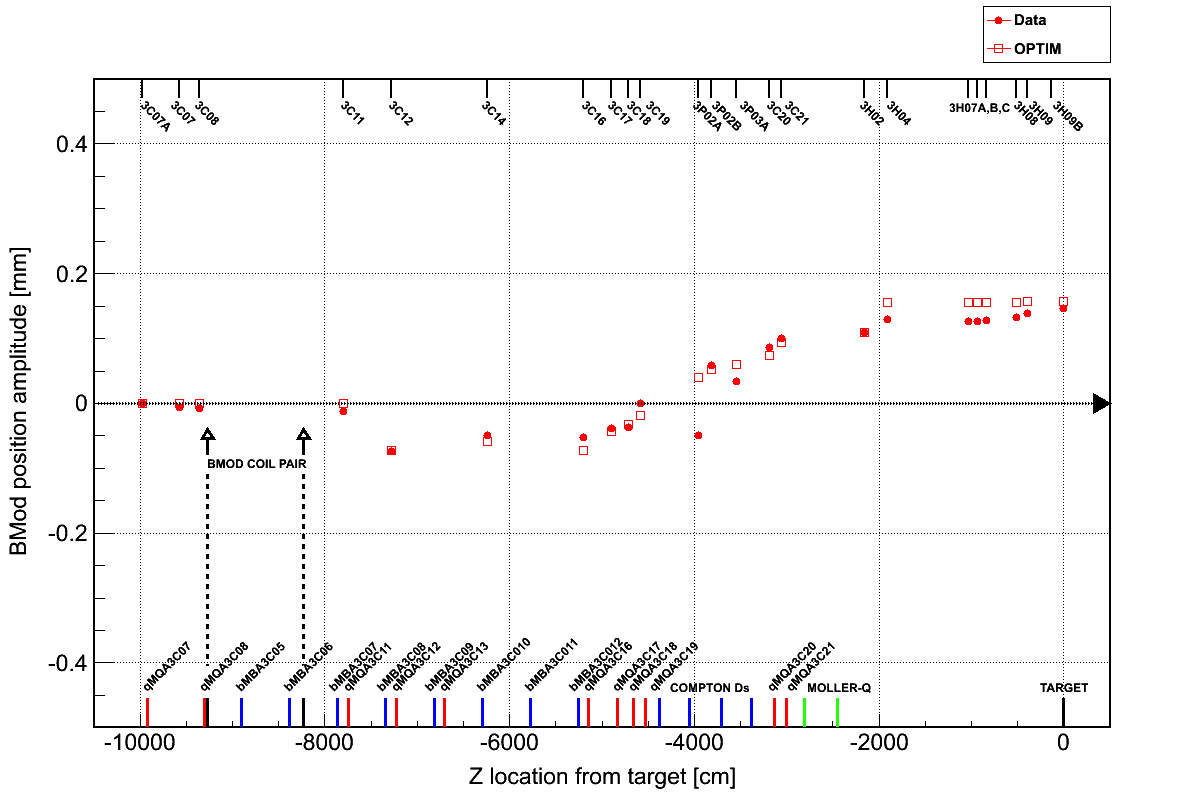}
  \caption{(Color online) Beam position monitor (BPM) amplitudes in response to X-modulation in the Hall-C beamline for one hour. Data (solid circles) and OPTIM (empty squares) \cite{optim}. The Z location of the BPMs are shown by short vertical lines at the top of the figure while the quadrupoles (red), dipoles (blue), moller quadrupoles (green), and other key components (black) are shown at the bottom.}
  \label{12000opticsXX}
\end{figure}

A typical BPM response is a sinusoid of amplitude $\sim$200 $\mu$m. Compared with natural beam jitter, this is an order of magnitude larger and has fewer correlations among the parameters, providing an independent way of measuring sensitivities. In FIGURE \ref{12000opticsXX}, data and OPTIM \citep{optim} are compared for X-positions of all the BPMs in the Hall-C beamline. 

\begin{figure}[h]
  \includegraphics[height=0.42\textheight]{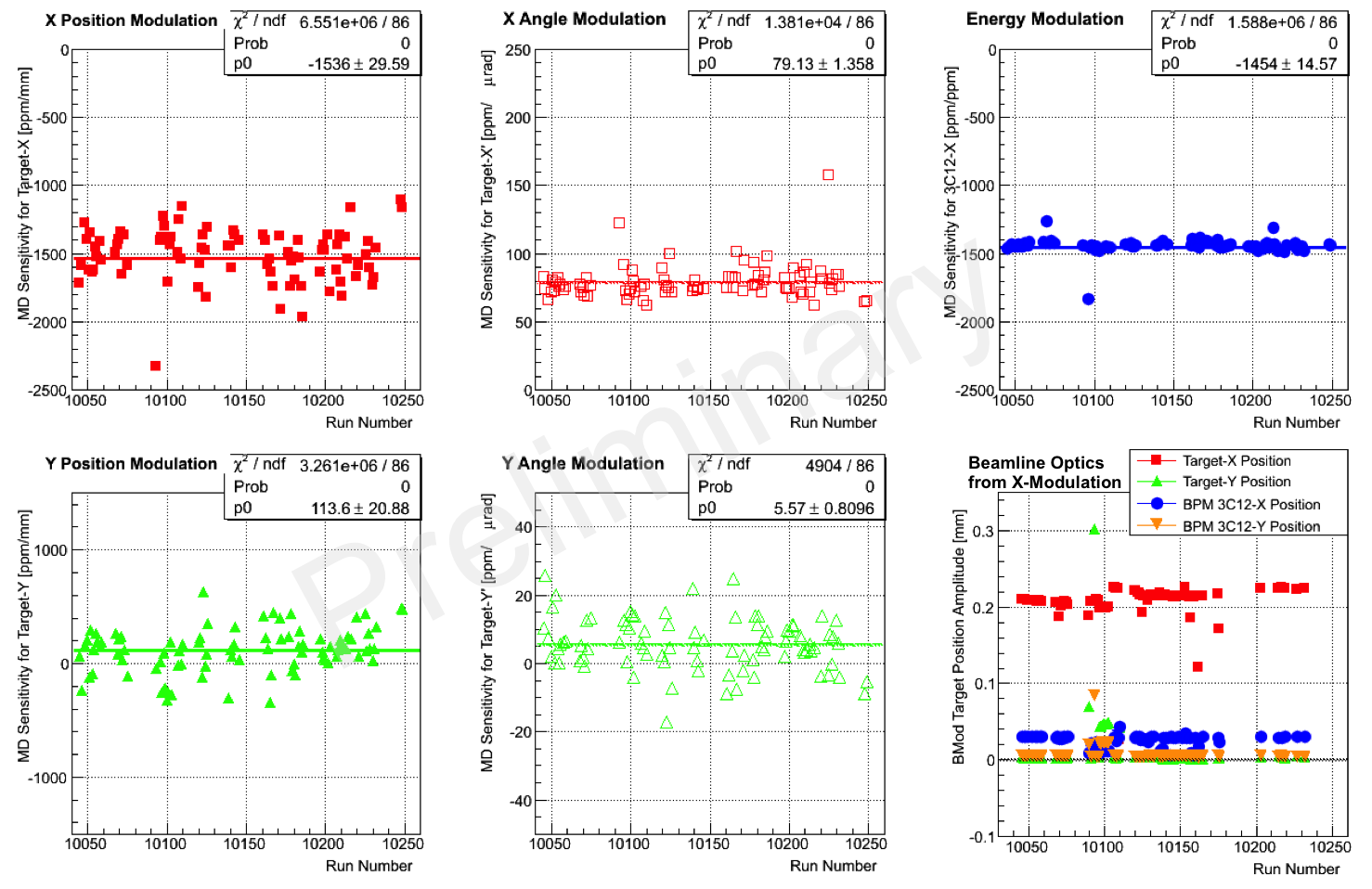}
  \caption{(Color online) Panel 1-5: Main detector sensitivities for  X (solid red square), X$^{\prime}$ (empty red square), E (solid blue square), Y (solid green triangle), and Y$^{\prime}$ (empty green triangle) are shown. Panel 6: The amplitude of target X (red square), Y (green triangle), 3C12 X (blue square), and Y (orange inverted triangle) responses for X-modulation.}
  \label{SensitivityOpticsXX}
\end{figure}

Detector responses correlated with modulation were used to extract preliminary sensitivities for the main detector and are shown in FIGURE \ref{SensitivityOpticsXX} (panel 1 through 5). The sensitivities for this period were fairly stable and successfully reduced the width of the regressed asymmetry.

The amplitude of the BPM response in front of the target and the middle of the arc (3C12) vs time are shown in FIGURE \ref{SensitivityOpticsXX} (panel 6). The optics were reasonably stable during this period except for several hours of X-Y coupling due to a quadrupole fault. 
From E modulation (not shown), we found that residual dispersion at the target was often large in both X and Y directions.

\section{Summary}
A pair of dipole, and a superconducting RF cavity were used to drive the beam along the Hall-C transport line to measure detector sensitivities for the Q$_{weak}$ experiment at Jefferson Lab. Preliminary sensitivities were fairly stable and successfully reduced the width of the measured asymmetry. The beam modulation system also has proven valuable for tracking changes in the optics such as dispersion at the target and X-Y coupling.

\bibliographystyle{aipproc}   


\bibliography{80_Nuruzzaman}

\IfFileExists{\jobname.bbl}{}
 {\typeout{}
  \typeout{******************************************}
  \typeout{** Please run "bibtex \jobname" to optain}
  \typeout{** the bibliography and then re-run LaTeX}
  \typeout{** twice to fix the references!}
  \typeout{******************************************}
  \typeout{}
 }

\end{document}